\documentclass[twocolumn,showpacs,preprntnumbers,amssymb,amsmath]{revtex4}
\usepackage{graphicx}
\usepackage{dcolumn}
\usepackage{bm}
\begin{document}
\preprint{CCOC-03-04 (Last Update:\today)}
\title{Map representation of the time-delayed system in the presence of Delay Time Modulation: 
an Application to the stability analysis} 
\author{Won-Ho Kye$^1$} 
\author{Muhan Choi$^1$}
\author{Tae-Yoon Kwon$^{1,2}$}
\author{Chil-Min Kim$^1$}
\affiliation{$^1$National Creative Research Initiative Center for Controlling Optical Chaos,
Pai-Chai University, Daejeon 302-735, Korea}
\author{Young-Jai Park$^2$}
\affiliation{$^2$Department of Physics, Sogang University, Seoul 121-742, Korea}

\begin{abstract}
We introduce the map representation of a time-delayed system in the presence of 
delay time modulation.
Based on this representation, we find the method by which to analyze the stability
of that kind of a system.
We apply this method to a coupled chaotic system and discuss 
the results in comparison to the system with a fixed delay time.
\end{abstract}

\pacs{05.45.Xt, 05.40.Pq}
\maketitle

In a real situation, time delay is inevitable, since the propagation speed of
an information signal is finite.
The effects of time delay have been investigated in various fields 
in a form of delay differential
equation \cite{Vol,TD_Cont,TD_Bio, TD_Laser, HyperChaos,NeuronSync}: 
\begin{equation}
\dot{x}=F({x}(t), {x}(t-\bar{\tau})), 
\end{equation}
where $\bar{\tau}$ is the delay time. 
Since the Volterra's predator-prey model \cite{Vol},
delay time has been considered such various forms as fixed \cite{TD_Bio,TD_Laser,HyperChaos,NeuronSync}, distributed\cite{Dist}, 
state-dependent \cite{Hale}, and time-dependent \cite{Hale,DTM,DTM_Sync} ones.
Up to now, the delay effects of these forms in dynamical systems
have been extensively studied not only in various fields of
physics\cite{TD_Laser}, biology \cite{TD_Bio}, and economy \cite{Hale} but also 
for the cases of application \cite{TD_Cont, TD_Comm, TD_Sync}.
The recent discovery \cite{NeuronSync} that the neural synchrony is enhanced by the discrete time delays
shows the delay feedbacks play a significant role in synchronization phenomenon.  
This discovery calls new scientific attention on the time-delay feedbacks and synchronization phenomenon. 

Meanwhile, in regard to the application of time-delayed systems to communication,
delay time modulation (DTM) \cite{DTM,DTM_Char,DTM_Sync,DTM_Death} 
driven by the signal of a independent system 
was introduced to erase the imprint of the fixed delay time. 
It was reported that the DTM increases the complexity
of the time-delayed system as well as prevents the system from being collapsed into
the simple manifold in reconstructed phase space, which was a serious drawback of the
time-delayed system to the application to communication.
It was also found that the synchronization in the presence of DTM can be established and 
the death phenomenon is enhanced, due to DTM\cite{DTM_Sync,DTM_Death}.

It was further shown that the time-delayed system can be represented by high dimensional coupled maps \cite{Map_Rep}
and the methods by which to understand the characteristics of the time delayed system with
discrete time delays have been developed \cite{TD_Cont}. 
Accordingly, the natural questions to be tackled have become "{\it how can we find the coupled-map representation of the
time-delayed system in the presence of DTM ?} " and " {\it how can we analyze 
the stability of that kind of system ?}".   
The answers on these two questions are quite essential to understand the synchronization and the enhancement 
of death phenomenon mentioned above with regard to DTM \cite{DTM_Sync, DTM_Death}.
In this paper, we newly introduce the coupled-map representation of a time-delayed system in the presence of
DTM, and develop a method which 
enables us to get the Lyapunov exponents of the system based on the representation.
We shall also show the DTM accelerates the transition to hyperchaos in a coupled time-delayed system 
by presenting the Lyapunov spectrum obtained by the method.

The delay differential equation of Eq. (1) can be rewritten by the discrete form: 
\begin{equation}
x_{n+1} = f(x_n, x_{n-\tau}),
\end{equation}
where $x_n$ is the value of $x$ at $n$-th time step such that $x_n=x(n \Delta t)$ 
and $x_{n-\tau}=x(n \Delta t -\tau \Delta t)$.
Here $f(x_n, x_{n-\tau})= x(n \Delta t)+F(x(n \Delta t) , x(n \Delta t-\tau \Delta t))\Delta t$. 
It is known that the above system is equivalent to the $(\tau+1)$ dimensional coupled maps \cite{Map_Rep}:
\begin{eqnarray}
x^0_{n+1}&=&  f(x^0_n, x^\tau_{n}),\nonumber\\
x^\tau_{n+1}&=& x^{\tau-1}_n,\nonumber\\
x^{\tau-1}_{n+1}&=& x^{\tau-2}_n,\nonumber\\
& \vdots &	\\
x^{1}_{n+1}&=&x^0_n, \nonumber
\end{eqnarray}
where the variables $x_n^1, \dots, x_n^{\tau}$ construct
the "echo type" feedback loop.
The above representation plays an important role in analyzing the 
time-delayed system with fixed delay times.
DTM \cite{DTM} consists of a modulation system, $\dot{y}= {G}({y})$ which generates the driving signal $y_n$,
and a scaling system, $\bar{\tau}=H({ y})$ which adjusts the amplitude of modulation.
Accordingly two additional equations are introduced:
\begin{eqnarray}
	{y}_{n+1} &=& g({ y}_n), \nonumber\\
	\tau&=&h({y}_n),
\end{eqnarray}
where $g(y_n)= y(n \Delta t) + G(y(n\Delta t))$ and $h(y_n)=H(y(n \Delta t))$.
The modulated delay time $\tau$ in Eq. (4) selects  
the length of delay time for a feedback variable $x_{n-\tau}$. 
As shown in Eq. (3) the feedback variable (say $x^\tau_n$) is not changed in the fixed delayed system.
However, in the time-dependent delayed system,  we can interpret that 
the feeding variable is selected according to the delay time (say $x^k_n$ with $k=0, 1, 2, ..., \tau_m$) in Eq. (3).    
Therefore we introduce the map representation as follows:
\begin{eqnarray}
{x}^0_{n+1}&=& f({x}^0_n, {x}^k_{n}),\nonumber\\
{x}^{\tau_m}_{n+1}&=& {x}^{\tau_m-1}_n,\nonumber\\
&\vdots&	\\
{x}^{1}_{n+1}&=&{x}^0_n, \nonumber
\end{eqnarray}
\begin{eqnarray}
k &= & h({ y}_n),  \\
{ y}_{n+1}&=& g({y}_n), \nonumber
\end{eqnarray}
where $\tau_m$ is the maximum of the amplitude of DTM and $k$ is an integer number 
which determines the variable to be fed into the systems. 
We have $R^{\tau_m+1} \times I \times R^{1}$ phase space. 
From Eq. (5) and (6), we get the Jacobian matrix as follows:
\begin{widetext}
\begin{equation}
{\bf J}=	\left[~
		\begin{matrix}
		\frac{\partial f(x^0_n, x^k_{n})}{\partial x_n^0} &	
		\frac{\partial f(x^0_n, x^k_{n})}{\partial x^k_n} \delta_k^{\tau_m} &
		\frac{\partial f(x^0_n, x^k_{n})}{\partial x^k_n}\delta_k^{\tau_m-1} & 
		\frac{\partial f(x^0_n, x^k_{n})}{\partial  x^k_n} \delta_{k}^{\tau_m-2} & \cdots& 
		\frac{\partial f(x^0_n, x^k_{n})}{\partial x^k_n}\delta_{k}^{1} &\frac{\partial f(x^0_n, x^k_{n})}{\partial k}  &0\\
		0& 0& 1 & 0 &\cdots& 0 & 0 & 0\\
		0& 0& 0 & 1 &\cdots& 0 & 0 & 0\\
		\cdots& \cdots &\cdots & \cdots &\cdots &\cdots &\cdots &\cdots \\
		0& 0& 0 & 0 &\cdots& 1 & 0 & 0\\
		1& 0& 0 & 0 &\cdots& 0 & 0 & 0\\
		0& 0& 0 & 0 &\cdots& 0 & 0 & \frac{\partial h(y_n)}{\partial y_n} \\
		0& 0& 0 & 0 &\cdots& 0 & 0 & \frac{\partial g(y_n)}{\partial y_n}
		\end{matrix}
	~\right], 
\end{equation}
\end{widetext}
where 
$\delta_k^l=1$ only if $k=l$ else $\delta_k^l=0$.
We introduce the ($\tau_m+3$) dimensional orthonormalized initial vectors such that: $\hat{\bf y}_1=(1, 0, 0, \dots )$,   
$\hat{\bf y}_2=(0, 1, 0, \dots)$, and $\dots$ which are evolved by $\hat{\bf y}_i^\prime = {\bf J} \hat{\bf y}_i$. 
By following the standard procedure \cite{Nayfeh}, 
we can find the Lyapunov spectrum of the time-delayed system in the presence of
DTM as follows:
\begin{equation}
	\lambda_i=\frac{1}{rT} \sum_{k=1}^r \ln N_i^k,
\end{equation}  
where $r$ is the number of orthonormalization within the chosen finite time interval $T$ and
$N_i^k$ is the norm of the $i$-{th} vector at $k$-{th} orthonormalization process.  



As an example of DTM, we consider the two coupled logistic maps as follows:
\begin{eqnarray}
	x_{n+1} &= & \gamma  \bar{x}_n^\tau(1-\bar{x}_n^\tau), \nonumber\\
	\tau & = &  [\Lambda y_n ], \nonumber\\		
	y_{n+1} &= &\gamma_0 y_n(1-y_n), 
\end{eqnarray}
Here $\bar{x}_n^\tau=(1-\alpha)x_n + \alpha( x_{n-\tau})$ and 
$\alpha$ is a  coupling strength.
We took $\gamma_0=4.0$ and $\Lambda=4.0$ and $[x]$ denotes the maximum of 
the integer which is less than $x$ and $\tau_m=\max ([\Lambda y_n])=3$.
According to the general representation of Eqs. (5) and (6),
the system described in Eq. (9) can be rewritten by the 6-dimensional coupled maps
as follows:
\begin{eqnarray}
x_{n+1}^0&=&\gamma \bar{x}_n^k (1- \bar{x}_n^k) \equiv F(\bar{x}_n^k), \nonumber\\
x_{n+1}^3&=&{x}_n^2, \nonumber\\
x_{n+1}^2&=&{x}_n^1, \nonumber\\
x_{n+1}^1&=&{x}_n^0, \\
 & & \nonumber \\
k&=& [\Lambda y_n], \nonumber \\
y_{n+1}&=&\gamma_0 y_n (1-y_n),  
\end{eqnarray}
where $\bar{x}_n^k=(1-\alpha)x_n^0 +\alpha x_n^k$.
The corresponding Jacobian matrix is given by:
\begin{widetext}
\begin{equation}
{\bf J}=	\left[~
		\begin{matrix}
		\gamma ((1-\alpha)+ \alpha \beta \delta_0^k) (1-2\bar{x}_n^k) &	
		\gamma \alpha \delta_3^k (1-2\bar{x}_n^k)  &	
		 \gamma \alpha \delta_2^k (1-2\bar{x}_n^k)&	
		 \gamma \alpha \delta_1^k (1-2\bar{x}_n^k)&	
		 F(\bar{x}_n^{k+1})-F(\bar{x}_n^k)& 0 \\	
		0& 0& 1 & 0 & 0 & 0\\
		0& 0& 0 & 1 & 0 &0 \\
		1& 0& 0 & 0 & 0 &0 \\
		0& 0& 0 & 0 & 0 &0 \\
		0& 0& 0 & 0 & 0 & \gamma_0 (1-2 y_n) \\
		\end{matrix}
	~\right], 
\end{equation}
\end{widetext}
where we have used the relation $\frac{\partial F(x^0_n, x^k_n)}{\partial k} =F(x^0_n, x^{k+1}_{n})-F(x^0_n, x^{k}_n)$
noting that $k$ is an integer \cite{ChainRule} and
Here $\frac{\partial [\Lambda y_n]}{\partial y_n}$ is not zero only at $0.25$, $0.50$, and $0.75$ which are
unstable periodic orbits of the driving logistic map (the second equation of Eq. (11)).
Therefore we put $\frac{\partial [\Lambda y_n]}{\partial y_n}=0$ to the above equation.

\begin{figure*}
\begin{center}
\rotatebox[origin=c]{0}{\includegraphics[width=13.0cm]{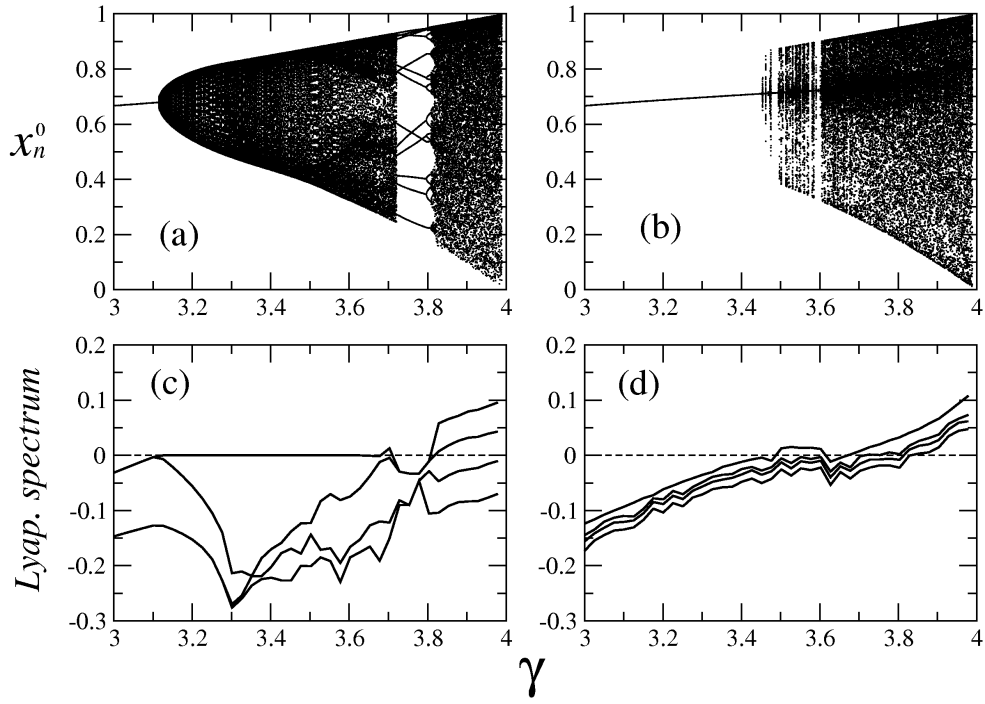}} 
\caption{The bifurcation diagrams and the Lyapunov spectra with $\alpha=0.7$. (a) and (c) show
 the bifurcation diagram and Lyapunov spectra at fixed delay time (only using Eq. (10) with $k=\tau_m$),
 (b) and (d) at DTM (using Eq. (10) and (11)). We have taken $r=5000$ orthonormalization 
process with the period of $T=70$ for each point (see Eq. (8)).}
\end{center}
\end{figure*}
To verify our method, we calculate the Lyapunov spectrum in the case of DTM as well as in the fixed delay case.
We can easily restore Eq. (10) to the fixed delay case by fixing $k=3$ without Eq. (11).
One can see the bifurcation diagram presented in Fig. 1 (a) and the corresponding
Lyapunov spectrum in Fig. 1 (c).
The quasi-periodic regime in $\gamma \in [3.1, 3.7]$ and the periodic window in $\gamma \in [3.7, 3.8]$
are exactly matched with the corresponding Lyapunov spectra. 
As one can see, the bifurcation diagram is different from that of the single logistic map,  
which is due to the delay feedback \cite{Map_Rep}.
In the case of fixed delay (Fig. 1 (a) and (c)), two largest Lyapunov exponents become positive above $\gamma=3.8$
through the quasi-periodic regime in $\gamma \in [3.1,  3.7]$. 
That is to say, the system develops to hyperchaos above that point.

The Lyapunov exponents in the case of DTM are presented in Fig. 1 (d). Actually we have two more  
Lyapunov exponents which are not plotted in the figure. 
One has a constant value, $\lambda_5=0.691\dots$, which describes the dynamics of isolated master system 
and the other has a relatively large negative value, $\lambda_6 \in [-0.5, -0.6]$. 
While in fixed delay time, $x_n^\tau$ is directly fed into the delayed system,
in DTM all delay variables $x_n^0, x_n^1, \dots x_n^\tau$ can contribute to 
the system depending on the state of the modulating signal $y_n$. 
For this reason, the largest Lypaunov exponents are very closer with each other (Fig. 1 (c)) than 
those of the fixed delay time (Fig. 1 (d)). 
The sum of positive Lyapunov exponents $\sum_{\lambda_i > 0} \lambda_i$ is the
entropy of the system which quantifies the complexity of the system \cite{Arba}.
For example, at $\gamma=3.95$ the sum of positive Lyapunov exponents 
is 0.262 in the case of DTM, while it is 0.128 in the case of fixed delay time. 
This type of behavior caused by DTM is fulfilled with the result of the recent report 
in which the entropy is increased depending on the property of the driving signal for DTM \cite{DTM_Char}.
 
Figure 2 (a) shows the time series at the intermittent regime ($\gamma=3.5$) and (b) shows the 
time series at the hyperchaotic regime ($\gamma=3.95$) when the delay is modulated.
The return maps are presented in Fig. 2 (c) and (d), which show a complex structure.
Comparing the correlation functions in  Fig. 2 (e) (with the fixed delay time) and (f) (with DTM), 
one is confirmed that the DTM reduces the length of correlation \cite{DTM_Sync}.  
We emphasize that all the procedure can be
applied to another system regardless of coupled maps or flows in principle
because in case of flow, the system also can be represented by the same form of 
coupled maps of Eqs. (5) and (6) (one just needs to note that the dimension of the phase space is determined by the
value of delay time $\tau$ and the chosen time step $\Delta t$ such that $\dim \sim \tau/\Delta t$). 

In conclusion, we have introduced the coupled-map representation of 
the time-delayed system in the presence of delay time modulation and 
found the method by which to analyze the stability of the system. 
We have applied this method to a coupled chaotic system and 
confirmed that the Lyapunov spectrum of the time-delayed system in the presence of DTM 
can be exactly determined which explains the bifurcation behavior.
Furthermore, we have observed that DTM erases the preference 
of the feedback variable, by scanning all delay variables during the evolution,
and that for this reason it leads Lyapunov exponents to be close to each other.   

The authors thank M.S. Kurdoglyan and K. V. Volodchenko for valuable discussions.
This work is supported by Creative Research Initiatives 
of the Korean Ministry of Science and Technology.

\begin{figure}
\begin{center}
\rotatebox[origin=c]{0}{\includegraphics[width=7.0cm]{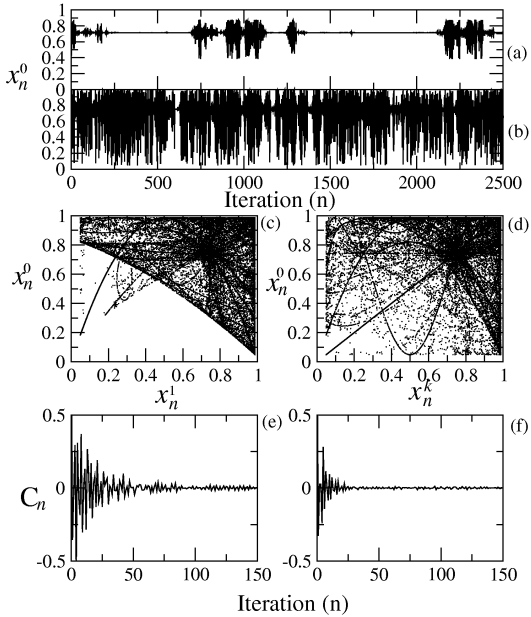}}
\caption{Temporal behaviors and retrun maps of delayed logistic map in the presence of DTM (Eq. (10) and (11)).
(a) Time series of $x_n^0$ at $\gamma=3.5$ (b) at $\gamma=3.95$; (c) return map $x_n^0$ versus $x_n^1$
(d) $x_n^0$ versus $x_n^k$; autocorrelation functions of $x_n^0$ as a function of iteration at $\gamma=3.95$
in cases of (e) fixed delay and (f) DTM.}
\end{center}
\end{figure}


\begin{thebibliography}{150}
\bibitem{ChainRule} Alternatively, using the chain rule, we can use the relation $\frac{\partial F(x^0_n, x^k_n)}{\partial k} =
			\frac{\partial F(x^0_n, x^k_n)}{\partial x^k_n}\frac{\partial x^k_n}{\partial k}=
			\frac{\partial F(x^0_n, x^k_n)}{\partial x^k_n}(x^{k+1}_n-x^k_n)$.
			We confirmed that the Lyapunov spectrum is the same within the error bound  of the order $\sim 10^{-4}$.
\end{thebibliography}
\end{document}